\documentclass[aps,epsfig,twocolumn]{revtex4}%
\usepackage{graphicx}
\usepackage{epsfig}
\usepackage{amsmath}
\usepackage{amsfonts}
\usepackage{amssymb}%
\begin{document}
\title{ Path Dependence of the Quark Nonlocal Condensate within the Instanton Model. }
\author{L.A.Trevisan$^{1,3}$, A. E. Dorokhov$^{2}$, Lauro Tomio$^{3}$,}
\affiliation{$^1$Departamento de Matem\'atica, Universidade Estadual de Ponta Grossa, Av.
Carlos Cavalcante, 4748, 84030-000, Ponta Grossa, PR, Brasil\\
$^2$Bogoliubov Laboratory of Theoretical Physics, JINR, Dubna 141980, Russia\\
$^{3}$Instituto de F\'{\i}sica Te\'{o}rica, Universidade Estadual Paulista,
Rua Pamplona 145, 01405-900, S\~{a}o Paulo, SP, Brasil}

\begin{abstract}
Within the instanton liquid model, we study the dependence of the gauge
invariant two--point quark correlator on the path used to perform the color
parallel transport between two points in the Euclidean space.

\end{abstract}
\maketitle

\section{Introduction}

The nonlocal vacuum condensates or vacuum correlators describe the
distribution of quarks and gluons in the non-perturbative QCD
vacuum~\cite{MihRad92}. Physically, it means that vacuum quarks and gluons can
flow through the vacuum with nonzero momentum. Nonlocal condensates play an
important role in hadron physics. In particular, they are of principle
importance in the study of the distribution functions of quarks and gluons in
QCD vacuum and in hadrons~\cite{MihRad92,DEM97,DoLT98}.

In general, the gauge invariant (and path independent) nonlocal quark-quark
correlator is defined by
\begin{equation}
\mathcal{M}(x,y)=\sum_{\mathcal{C}}e^{-L[\mathcal{C}]}\mathcal{M}
^{\mathcal{C}}(x,y),\label{corr}
\end{equation}
where {\small
\begin{align}
&  \mathcal{M}^{\mathcal{C}}(x,y)=({1}/{N_{c}})\times\label{string}\\
&  \langle0|:\mathrm{Tr}\left\{  \mathrm{P}\exp\left(  -ig\int_{~\mathcal{C}
\left(  x,y\right)  }dz^{\mu}A_{\mu}^{a}(z)T^{a}\right)  q(x)\overline
{q}(y)\right\}  :|0\rangle\nonumber
\end{align}
} is the quark correlator with the Schwinger phase factor path-ordered along
arbitrary path $\mathcal{C}\left(  x,y\right)  $ from $x$ to $y$. The
P-exponential ensures the parallel transport of color from one point to
another. $T^{a}$ are the matrices of the algebra of the color group SU(N$_{c}
$) in the fundamental representation. The trace is taken over the color
indices and $\exp{\{-L[\mathcal{C]\}}}$ is the dynamical weight of each path configuration.

\begin{figure}[ptb]
\centering \includegraphics[width=2.5in]{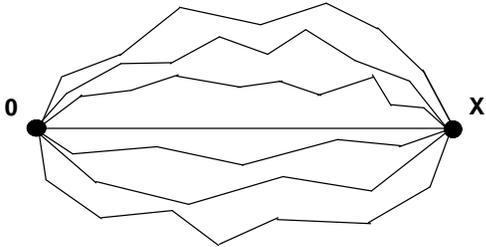}\caption{Schematic
representation of the quark correlator with arbitrary paths between the
quarks, which are in the positions 0 and $x$.}
\end{figure}

To evaluate Eq.~(\ref{corr}), one has to calculate the nonperturbative
expectation value of the Wilson line operator for an arbitrary path
$\mathcal{C}$ and perform resummation over all possible path configurations,
pictorically represented in Fig. 1. Both tasks are extremely difficult and
cannot be performed in full manner at the current stage. In the field theory
there is a special class of observables for which the straight-line paths are
dominated and the path integrations can be performed exactly. Propagation of
an energetic quark through a cloud of soft gluons, and the behaviour of light
quark in the field of infinitely heavy quark are some of examples. Such an
approximation is called the straight-line path approximation~\cite{Barb65}.

The present note is devoted to study the path dependence of the quark-quark
correlator in the instanton liquid model.  In sections II and III, we consider
the quark correlator for the straight-line and broken line paths. In section
IV we present our results and conclusions.

\section{Straight-line path approximation}

The gauge invariant correlator $\mathcal{M}^{\mathcal{C}}(x,y)$ in
(\ref{corr}) depends on the choice of the path $\mathcal{C}$. Usually, the
path $\mathcal{C}$ appearing in Eq.~(\ref{string}) is assumed to be a straight
line connecting the points $x$ and $y$. With the path, that is a straight line
connecting the points $0$ and $x$, the Schwinger string in (\ref{string})
reads:
\begin{equation}
S(0,x)=\mathrm{P}\exp\left(  -ig\int_{0}^{1}dt\,x^{\mu}A_{\mu}(xt)\right)
~.\label{line}
\end{equation}
Then, in the Euclidean region, the translational, O(4) and parity symmetries
require the correlator (\ref{string}) to be of the following form:
\begin{equation}
\mathcal{M}^{(\mathcal{C})}(0,x)=\langle0|:\overline{q}(0)q(0):|0\rangle
\mathcal{Q}^{(\mathcal{C})}(x^{2})~,\label{param}
\end{equation}
where $\mathcal{Q}$ is an invariant function of $x^{2}$. In principle, every
choice of a path $\mathcal{C}_{[0,x]}$ connecting the two points $0$ and $x$
in the expression (\ref{string}) for the Schwinger string operator will
generate a different quark correlator, that we have denoted in
Eq.~(\ref{param}) as $\mathcal{Q}^{(\mathcal{C})}(x)$ [$\mathcal{Q}
^{(\mathcal{C})}(0)=1$].

Recently, the quark nonlocal condensate, with the straight--line parallel
transport, has been evaluated on the lattice \cite{Delia99} and
semi--classically in the instanton dilute--liquid model \cite{DEM97}. The
instanton model of the QCD vacuum is a realistic one in prediction of the
behaviour of the quark and gluon correlation functions
\cite{DEM97,DoLT98,DEM99}. For the straight line path the averaging over the
instanton vacuum gives in the single--instanton approximation the result
\cite{DEM97,DoLT98}
\begin{align}
\mathcal{Q}_{I}(x^{2}) &  =\frac{8\rho^{2}}{\pi}\int_{0}^{\infty}drr^{2}
\int_{-\infty}^{\infty}\ dt\times\nonumber\\
&  \times\frac{\cos\left[  \frac{r}{R}[\arctan(\frac{t+|x|}{R})-\arctan
(\frac{t}{R})]\right]  }{[R^{2}+t^{2}]^{3/2}[R^{2}+(t+|x|)^{2}]^{3/2}},
\end{align}
\noindent where $\rho$ is the average instanton size and $R=\sqrt{\rho
^{2}+r^{2}}$.

In the next, we discuss the path dependence of the correlators and present the
formalism that we are considering.

\section{Path dependence of the nonlocal quark condensate}

The nonlocal quark condensate with straight--line path has a direct physical
interpretation in the heavy quark theory of heavy--light mesons as it
describes the propagation of a light quark in the color field of an infinitely
heavy quark. In the light quark sector in principle one needs to know the path
dependence of the correlator and average over all possibilities, as shown in
Fig.~1. Moreover, knowledge of the path dependence is also important in
modeling the higher dimensional nonlocal condensates used in QCD sum rules for
hadron wave functions \cite{BM95}.

No analysis exists of what happens for a different choice of the path
$\mathcal{C}$ other than the straight line, except recent very intersting
results on the path dependence of the gauge invariant gluon field strength
correlator studied in numerical simulations on a lattice in
Ref.~\cite{DiGiacomo:2002mq}. Here, we report our results on the path
dependence of the gauge--invariant quark two-point correlator within the
instanton vacuum model for arbitrary paths made of two joined straight lines
with cusp (polygon path). The broken-line quark condensate has the general
structure (see, \textit{e.i.} \cite{Mikh92,Grozin94})

\begin{align}
\mathcal{M}^{\mathcal{C}}(x,y)  &  =\left\langle \overline{q}q\right\rangle
\left[  \mathcal{Q}_{S}^{(\mathcal{C})}(x,y)+\frac{\lambda_{q}^{2}}
{24}\mathcal{Q}_{T}^{(\mathcal{C})}(x,y)[\hat{x},\hat{y}]\right. \nonumber\\
&  -\left.  \frac{i}{4}\left(  \mathcal{Q}_{V}^{(\mathcal{C})}(x,y)\hat
{x}-\mathcal{Q}_{V}^{(\mathcal{C})}(y,x)\hat{y}\right)  \right]  ,
\label{BrokQ}
\end{align}
\noindent where $\lambda_{q}^{2}$ is an average quark virtuality in the QCD vacuum.

In order to calculate the matrix element (\ref{string}) we use the instanton
liquid model. In the single instanton approximation the vacuum expectation
value is found by taking the classical instanton solution and quark zero modes
developed in its field and averaging over instanton collective variables: its
size, color orientation and position in space. Then one gets the gauge
invariant expression
\begin{align}
&  Q_{I,i}^{(\mathcal{C})}(x)=\int dD\left(  \rho\right)  \int dU\int
dz_{0}^{4}\frac{1}{N_{c}}Tr\{\overline{q}_{0}(x-z_{0})\Gamma_{i}\nonumber\\
&  \times P\exp\left[  -ig\int_{x}^{y}dz_{\mu}A_{I\mu}^{a}\left(
z-z_{0}\right)  \frac{\tau^{a}}{2}\right]  q_{0}\left(  y-z_{0}\right)
\}.\label{QIx}
\end{align}
\noindent For explicite calcualtions we choose the regular gauge for the
instanton field
\begin{equation}
A_{I\mu}^{a}\left(  z\right)  =\frac{1}{g}\eta_{\mu\nu}^{a}z_{\nu}\Phi\left(
z\right)  ,\qquad\Phi\left(  z\right)  =\frac{2}{\rho^{2}+z^{2}},\label{Ainst}
\end{equation}
and corresponding quark zero mode
\begin{equation}
q_{0}\left(  z\right)  =\frac{\rho}{\pi}\frac{1}{\left(  \rho^{2}
+z^{2}\right)  ^{3/2}}\chi,\qquad\overline{\chi}\chi=2.\label{Qzm}
\end{equation}
In Eq.~(\ref{Ainst}), $\eta_{\mu\nu}^{a}$\ is the antisymmetric 't Hooft
symbol. The broken--line path is parameterized in the form
\begin{equation}
\mathcal{C}_{[y,0,x]}=\left\{
\begin{array}
[c]{c}
y_{\mu}\lambda,\qquad-1\leq\lambda\leq0,\\
x_{\mu}\lambda^{\prime},\qquad0\leq\lambda^{\prime}\leq1,
\end{array}
\right.
\end{equation}
\noindent with the cusp angle $\alpha$ defined as
\[
\cos(\alpha)=\frac{y\cdot x}{|x||y|}.
\]
Note, that in the zero mode approximation, $\mathcal{Q}_{V}^{(\mathcal{C}
)}(x,y)$ is absent due to its odd chirality structure. Let us consider the
scalar structure, $\mathcal{Q}_{S}^{(\mathcal{C})}(x,y)$, in Eq.~(\ref{BrokQ}
). The final result, which is gauge and translation invariant, is given by
\begin{align}
&  Q_{S\mathrm{,I}}^{(\mathcal{C})}(y,x)=n_{c}\int dz_{0}^{4}\overline{q}
_{0}(y-z_{0})q_{0}\left(  x-z_{0}\right)  \times\nonumber\\
&  \times\left\{  \cos\left[  \alpha_{y}\right]  \cos\left[  \alpha
_{x}\right]  +\left(  n_{y},n_{x}\right)  \sin\left[  \alpha_{y}\right]
\sin\left[  \alpha_{x}\right]  \right\}  ,\label{Qbroken}
\end{align}
where the corresponding phase factors $\alpha_{x,y}$ and unit vectors in color
space $n_{x,y}^{a}$ $\left(  n_{x,y}^{2}=1\right)  $ are given by
\begin{align}
\alpha_{u} &  =\frac{1}{2}\sqrt{u^{2}z_{0}^{2}-\left(  uz_{0}\right)  ^{2}
}\int_{0}^{1}d\lambda\Phi\left(  u\lambda-z_{0}\right)  ,\\
n_{u}^{a} &  =\frac{u_{\mu}\eta_{\mu\nu}^{a}z_{0\nu}}{\sqrt{u^{2}z_{0}
^{2}-\left(  uz_{0}\right)  ^{2}}},
\end{align}
with $u=$ $x$ or $y$. The scalar product in color space is
\begin{equation}
\left(  n_{y},n_{x}\right)  =\frac{\left(  y,x\right)  z_{0}^{2}-\left(
z_{0},y\right)  \left(  x,z_{0}\right)  }{\sqrt{y^{2}z_{0}^{2}-\left(
yz_{0}\right)  ^{2}}\sqrt{x^{2}z_{0}^{2}-\left(  xz_{0}\right)  ^{2}}}.
\end{equation}

In the next section, we present and discuss the results of our calculations.

\begin{figure}[ptb]
.\vskip -3cm \centering \includegraphics[width=3in]{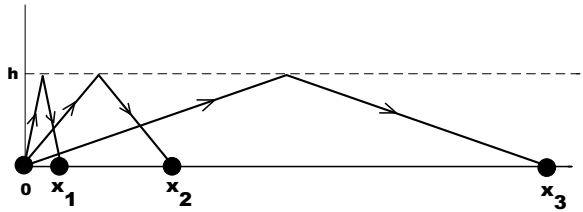}
\vskip -3cm\caption{ Different paths between quarks that are at fixed
deflection from the straight line. The position of the cusp is at equal
distance from each quark.}
\end{figure}

\section{Results and conclusions}

We have explored the dependence on the path by calculating the correlator,
considering various different paths, as shown in Figs. 2 and 4. The
corresponding results are presented in Figs. 3 and 5. All distances are given
in units of the instanton size $\rho$, i.e., for $\rho=1$.

\begin{figure}[ptb]
\centering \includegraphics[width=3.5in]{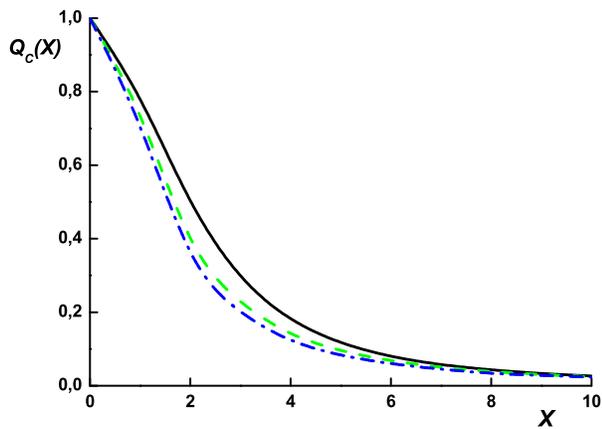}\caption{ Quark
correlator for different paths shown in Fig. 2. The solid line is for the
straight-line path ($h=0$); dashed line is for the broken-line path, with
$h=3$; and dot-dashed line is for $h=6$. All the distances ($x$, $h$) are in
units of the instanton size $\rho$.}
\end{figure}

\begin{figure}[ptb]
\centering \includegraphics[width=3in]{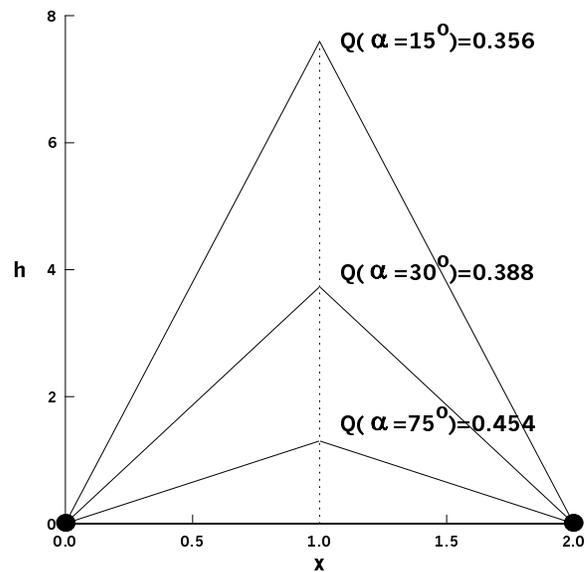} \vspace{-1cm}\caption{
Different paths between quarks that are at fixed distance $x=2$, considering
the same distance of the cusp from each quark. Indicated inside the figure we
have three representative cases, with the corresponding values of the
correlator. }
\label{f4}
\end{figure}

\begin{figure}[ptb]
\centering \includegraphics[width=3.5in]{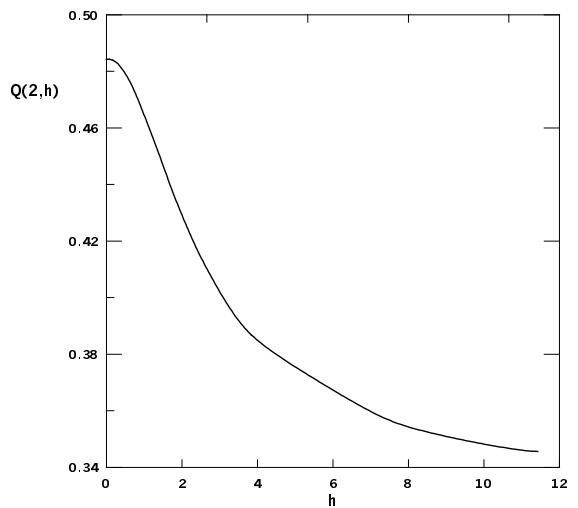} \vspace{-2.5cm}\caption{
Quark correlator for different paths shown in Fig.\ref{f4}, as a function of
the cusp distance. The quarks are at fixed distance $x=2$ between them, and at
the same distances from each cusp. }
\end{figure}

In Fig.~2, we show a set of paths with fixed \textit{transverse} size $h$
parameterizing the deviation from the straight--line path. A similar set of
paths has been considered in lattice calculations \cite{DiGiacomo:2002mq} for
the gluon field-strength correlator. We have chosen the position of the
\textit{cusp} in the broken-line path in Fig. 2 in the middle of the line
$(0,x)$. All quark correlators are normalized at zero distance, $x=0$,
independently on the form of closed path. All curves in Fig.~3 have the same
large $x$ asymptotics. This is because at large distances the relative
deflection $h$ becomes very small ($h<<x$) and the path becomes almost
indistinguished from the straight--line path.

The same considerations also apply to the \textit{sets} of paths shown in Fig.
4, for a fixed distance between quarks and equal distance of each cusp from
the quarks. This set of paths, with fixed equal distances from the cusps to
the quarks, shows the maximal effects related to the deflection of the
broken-line path from the straight-line one. The results for the correlator
$Q(x,h)$ are shown in Fig.~3, versus the distance $x$; and, in Fig. 5, versus
the distances $h$, related to the angle $\alpha$.

Finally, we observe that the main result is the relatively strong dependence
of the correlators on the shape of the path (Fig. 5), which can particularly
be more relevant for the light quark propagation. The correlator corresponding
to the straight--line path, given in Eq. (\ref{line}), has the largest signal
(for every distance $h$), compared with other choices of the path (Fig. 3).
Every deformation from the straight--line path produces a smaller value for
the correlator.

\section*{Acknowledgments}

\vspace{2mm} Our thanks to Funda\c{c}\~{a}o de Amparo \`{a} Pesquisa do Estado
de S\~{a}o Paulo (FAPESP), that makes possible this collaboration. L.T. also
thanks the partial support received from Conselho Nacional de Desenvolvimento
Cient\'{\i}fico e Tecnol\'{o}gico do Brasil. AED thanks for partial support
from RFBR (Grants nos. 02-02-16194, 03-02-17291, 04-02-16445), INTAS (Grant
no. 00-00-366).

\end{document}